\documentclass[a4paper,english,preprint,aps,prl]{revtex4-2}
\usepackage{amsmath}
\usepackage{amsmath, amssymb}
\usepackage{graphicx}
\usepackage{hyperref}
\usepackage{amsfonts}
\usepackage{ragged2e}
\usepackage{babel}
\justifying



\begin{document}


\title{Wave packet motion in a quantized electromagnetic field: Analytic results}

\author{Szabolcs Hack$^{1,2,*}$}
\email{szabolcs.hack@eli-alps.hu}

\author{B\'{e}la G\'{a}bor Pusztai$^{3,4,*}$, Krisztina Jo\'{o}s$^{2,5}$, S\'{a}ndor Varr\'{o}$^{1}$, Attila Czirj\'{a}k$^{1,2}$ and P\'{e}ter F\"{o}ldi$^{1,2}$}

\affiliation{$^1$ ELI ALPS, The Extreme Light Infrastructure ERIC, Wolfgang Sandner u. 3., 6728 Szeged, Hungary}%

\affiliation{$^2$Department of Theoretical Physics, University of Szeged, Tisza Lajos k\"{o}r%
\'{u}t 84, H-6720 Szeged, Hungary}

\affiliation{$^3$Bolyai Institute, University of Szeged, Aradi vértanúk tere 1, H-6720 Szeged, Hungary}

\affiliation{$^4$Department of Basic Sciences, Faculty of Mechanical Engineering and Automation, John von Neumann University, Izs\'aki \'ut 10, H-6000 Kecskem\'et, Hungary}

\affiliation{$^5$Faculty of Physics, Babe\c{s}-Bolyai University,
Str.\ Mihail\ Kog\u{a}lniceanu 1, RO-400084 Cluj-Napoca, Romania}

\affiliation{$^*$These authors contributed equally to this work.}


\begin{abstract}

We investigate the dynamics of a charged particle interacting with a multimode quantized electromagnetic field and obtain an analytic solution for the full electron--field system. This framework enables the calculation of position expectation values and uncertainties for arbitrary wave packets and field states, allowing us to identify quantum corrections to the corresponding classical motion. While the corrections to the position expectation value are weak and largely insensitive to the quantum state of the field, the wave packet broadening exhibits a pronounced dependence on the field state. In particular, the quantum uncertainty of the radiation is directly imprinted onto the spatial uncertainty of the particle. We illustrate these effects for Gaussian wave packets interacting with coherent, Fock, and squeezed states, including bright squeezed vacuum. The interaction with a finite-duration laser pulse is also analyzed as a multimode example. Our results provide a transparent analytic route toward understanding how quantum fluctuations of light influence electron dynamics in strong-field settings.

\end{abstract}

\maketitle

\section{Introduction}

The quantum electrodynamical description of light--matter interaction has a long and distinguished history, in which the interaction of electrons with the quantized radiation field represents one of the most fundamental models (see e.g.~Ref.~\cite{Spohn2004} for an overview). Beyond its conceptual importance, this problem has recently gained renewed relevance due to the significance of field quantization in strong-field phenomena. The process of high-order harmonic generation (HHG) \cite{McPherson87,Ferray88} plays a central role in this field, mainly due to the generation of attosecond pulses \cite{FarkasToth1992_Attopulses,Krausz2024}. However, the well established traditional HHG models \cite{Corkum1993_Threestep,Lewenstein1994_HHGmodel} treat the exciting laser pulse as a classical, time-dependent external field. Despite their remarkable success, these approaches cannot account for effects that originate from the quantum optical structure of the excitation. Let us note that electrons liberated from atomic cores propagate predominantly under the influence of the exciting laser field (strong-field approximation), in which case our forthcoming results are directly applicable.

The interaction of electrons with a quantized electromagnetic field was already discussed in the mid-20th century: in Ref.~\cite{Smith1946}, perturbation theory was applied to describe the field inside a conducting enclosure, while in Ref.~\cite{Gabor1950} -- using the example of an electron traversing a waveguide -- the relevance of field quantization from the viewpoint of information theory was pointed out.
The first non-perturbative treatment of HHG with quantized fields dates back to the early 1980s \cite{BergouVarro1981_Nonrel}, but clear experimental signatures of quantum-optical effects in HHG have been obtained only much more recently \cite{Tsatrafyllis2017,T19}. Since then, additional quantum-optical aspects of HHG have been revealed, both theoretically \cite{Stammer2023_QEDHHG,Lewenstein2021,RiveraDean2024} and experimentally \cite{Theidel2024_QOptHHG,Theidel2025_DisplacedSqueezed}, demonstrating that nonclassical states of light can indeed be generated in strong-field processes. Apart from free-field, laser-induced effects, quantum correlations also emerge when free electrons interact with integrated waveguides \cite{Arend2025}.
The emergence of HHG spectra as expectation values of the photon numbers in the harmonic modes was shown in Ref.~\cite{GCVF2016}, a phase-space description of the problem was given in Ref.~\cite{GVMF20}, and a general parametric model of the process was introduced in Ref.~\cite{GVP24}.

Bright squeezed vacuum (BSV) has attracted particular attention as a possible efficient driver of strong-field phenomena. Recent works demonstrated its role in electron dynamics \cite{EvenTzur2024_BSVmotion}, in extending the harmonic cutoff \cite{EvenTzur2024_HHGsqueezed}, and in solid-state HHG \cite{Rasputnyi2024_BSVHHG}. Experimental results confirmed the possibility of driving HHG by BSV states \cite{Heimerl2025_BSVphotoemission}. The temporal structure of single BSV shots has recently been measured using spectral interferometry with a coherent-state reference pulse \cite{Kern2026}. In the present context, the recent numerical results \cite{EvenTzur2024_BSVmotion} are especially relevant, as they show wave packet width oscillations in BSV that are in accord with the analytic results we present below.

Our work strongly builds on earlier results \cite{BergouVarro1981_Nonrel,Varro_NJP_2008,Varro2010,Varro2021_Photonics}, where the problem of an electron represented by a single plane wave interacting with a quantized mode was solved and analyzed. Additionally, the description below (focusing on wave packet dynamics) is complementary to that of Refs.~\cite{Andrianov2024PRA,Andrianov2025PRA}, in which the authors have shown that the interaction with free electrons affects the quantum state of the field, leading to the formation of nonclassical field states. 

In the following, in Section~2 we introduce the model we consider. The analytic solution of the coupled electron--field dynamics is presented in Sections~3 and 4, while representative examples are given in Section~5. Conclusions are drawn in Section~6.

\section{Model}
In the following, we consider a charged particle (for the sake of simplicity, we will call it an electron) that is interacting with the radiation field. We focus on the nonrelativistic regime and write the corresponding Hamiltonian as
\begin{equation}\label{Ham_original}
H=H_e \otimes \mathbf{1}+\mathbf{1}\otimes H_f + H_{ef},
\end{equation}
where $\mathbf{1}$ stands for the identity operator. The first factor in the tensor product corresponds to the electron, while the second one to the field, which is itself a tensor product describing different modes. The operators $H_e$ and $H_f$ denote the free Hamiltonians of the electron and the field, respectively, while $H_{ef}$ describes the interaction. When it causes no confusion, the explicit tensor product notation will be omitted in the rest of the paper.

Throughout this work, dipole approximation will be used, i.e., the spatial dependence of the field will be neglected. This naturally raises limitations concerning the size of the wave packet, and it is valid only if the contribution of short-wavelength modes to the dynamics is negligible. This condition, as we shall see, holds for excitations e.g.\ in the infrared or even the visible regime. It is worth emphasizing that the dipole approximation — besides leading to analytic solutions — also allows us to investigate effects that are related to the quantized nature of the radiation field in a very transparent way. Since we are to focus on the real-space properties of the wave packets (position expectation value and uncertainty), the assumption of negligible spatial dependence of the field means investigating quantum effects in their purest form.

For the sake of simplicity, linear polarization will be considered. In this case, it is sufficient to focus on electron motion along the direction of polarization (to be denoted by $x$). Thus we can write \cite{Pauli1938}
\begin{equation}
H=\frac{1}{2m}\left(P-eA\right)^2 + \sum_n \hbar \omega_n\left( a_n^\dagger a_n + \frac{1}{2}\right),
\end{equation}
where $P$ is the $x$ component of the momentum operator, and 
\begin{equation}
A=A_x=\sum_n \mathcal{A}_n (a^\dagger_n+a_n)
\end{equation}
denotes the corresponding component of the vector potential, with $a^\dagger_n$ and $a_n$ being the creation and annihilation operators of the $n$th mode, the frequency of which is $\omega_n.$ Note that the assumption of a discrete mode structure (i.e., using a sum instead of an integral over the modes) is not crucial in the following; it merely simplifies the notation. With this convention we have $\mathcal{A}_n=\sqrt{\frac{\hbar}{2 \epsilon_0 \omega_n V}},$ where the quantization volume $V$ appears explicitly.

By expanding the square in $H$ above, we obtain a term proportional to $A^2$. A Bogoliubov transformation \cite{Bogoliubov1947} — which, within the dipole approximation, is equivalent to a gauge transformation — can be used to eliminate this term, leading to $H_{ef}$ that is linear in $A.$ See Ref.~\cite{Andrianov2024PRA} for the application of the Bogoliubov transformation in this context. Also note that the term proportional to $A^2$ disappears in the case of circular polarization \cite{Varro_NJP_2008}. Having that in mind, our starting interaction Hamiltonian in the following will be
\begin{equation}\label{Hef}
H_{ef}=-\frac{1}{m}ePA.
\end{equation}
Clearly, the other two terms in $H$ are given by:
\begin{equation}\label{He}
H_{e}=\frac{P^2}{2m},
\end{equation}
\begin{equation}\label{Hf}
H_f=\sum_n \hbar \omega_n\left( a_n^\dagger a_n + \frac{1}{2}\right).
\end{equation}

It is important to emphasize that $H=H_e+H_f+H_{ef}$ has no explicit time dependence; in other words, there is no ``external excitation" as a time-dependent classical driving field. However, in order to be able to compare the results to be presented with experimental findings, it is important to see how the expectation value of the linearly polarized electric field $E=\sum_n i \mathcal{A}_n \omega_n (a_n-a^\dagger_n)$ can be calculated. To this end, let $|\phi_{f}\rangle$ denote the initial state of the multimode field. In the absence of the electron ($H_e=H_{ef}=0$), the Heisenberg equations of motion for $E$ can be solved easily. In this case we have $\langle E \rangle_0(t)=\langle \phi_{f}\vert E_0^{H} (t)\vert \phi_{f}\rangle,$ where the indices 0 and $H$ refer to the free-field case and the Heisenberg picture, respectively, and  $E_0^{H} (t)=\sum_n i \mathcal{A}_n \omega_n (a_n e^{-i\omega_n t}-a^\dagger_n e^{i\omega_n t}).$ It is clearly $\langle E \rangle_0(t)$ that has to be identical to the ``waveform" of the electric field that can be measured in an experiment:
\begin{equation}\label{Qcl}
E_{cl}(t)=\langle E \rangle_0(t).
\end{equation}
More details on this correspondence will be given in subsec.~5.2.

\section{Analytic solution of the time-dependent Schr\"{o}dinger equation}
The dynamics of the coupled field-electron system is governed by the time-dependent Schr\"odinger equation
\begin{equation}\label{Schr}
i\hbar \frac{\partial}{\partial t}\vert\Psi(t)\rangle=H\vert\Psi(t)\rangle
\end{equation}
with $H$ being given by Eqs.~(\ref{Ham_original}),(\ref{Hef}),(\ref{He}) and (\ref{Hf}). Considering an interaction with a pulsed (i.e., finite duration) field, we can assume that the states of the electron and the field initially (``before the interaction", taken as $t=0$) are not correlated. They can be represented by a tensor product:
\begin{equation}\label{initial}
\vert\Phi(t=0)\rangle=\vert\phi_e\rangle\otimes \vert \phi_f\rangle,
\end{equation}
where $\langle \phi_e\vert\phi_e\rangle=\langle \phi_f\vert\phi_f\rangle=1.$ We assume that initially $\vert \phi_f\rangle$ is itself a tensor product state, i.e.,
\begin{equation}\label{initialfield}
\vert \phi_f\rangle=\prod_n^{\otimes} \vert \phi_{f,n}\rangle.
\end{equation}

In practice, a high-frequency cutoff may be necessary (see e.g.~\cite{CohenTannoudji1989}), which is consistent with using the dipole approximation. Focusing on an experimental situation when a finite-duration laser pulse (given by $E_{cl}(t)$) interacts with the electron, the modes outside the spectral range of this excitation are initially in vacuum state. Additionally, the expectation value of the free-field electric field must satisfy Eq.~(\ref{Qcl}).

As an example, let us consider a tensor product of coherent states
\begin{equation}\label{cohprod}
\vert \phi_f\rangle=\vert \boldsymbol{\alpha}\rangle=\prod_n^{\otimes} \vert \alpha_n\rangle=\mathbf{D}(\boldsymbol{\alpha})|\mathbf{0}\rangle=\prod_n^{\otimes} D_n(\alpha_n)\vert 0\rangle_n,
\end{equation}
where a single mode displacement operator is given by
\begin{equation}\label{Dsingle}
D(\alpha)=e^{\alpha a^\dagger-\alpha^* a}.
\end{equation}
Since $a\vert \alpha\rangle=\alpha \vert \alpha\rangle,$ we have
\begin{equation}\label{cohfield}
\langle E \rangle_0(t)=\sum_n i \mathcal{A}_n \omega_n \left(\alpha_n e^{-i\omega_n t}-\alpha^*_n e^{i\omega_n t}\right).
\end{equation}
In view of this, condition (\ref{Qcl}) can be used to determine the labels $\alpha_n$ at $t=0,$ see subsec.~5.2.

Considering the initial state of the electron, we assume a normalized wave packet that corresponds to a wavefunction $\phi_e(x).$ More formally,
\begin{equation}\label{estate0}
\vert\phi_e\rangle=\int \phi_e(x) \vert x\rangle dx=\int \widetilde{\phi_e}(p) \vert p\rangle dp,
\end{equation}
where $X\vert x\rangle=x\vert x \rangle,$ $P\vert p\rangle=p\vert p \rangle$ and $\widetilde{\phi_e}(p)=\frac{1}{\sqrt{2\pi\hbar}}\int \exp(-i px/\hbar) \phi_e(x) dx$.

\subsection{Diagonalizing the Hamiltonian}

The interaction term in the Hamiltonian
\begin{equation}\label{Hamstart}
H=\frac{P^2}{2m}+\sum_n \hbar \omega_n\left( a_n^\dagger a_n + \frac{1}{2}\right) - \frac{1}{m}ePA
\end{equation}
can be eliminated by using the idea introduced in Ref.~\cite{BlochNordsieck1937}. To this end, we employ
a generalized displacement operator that acts nontrivially on all factors of the Hilbert space of the complete system:
\begin{equation} \label{Dhat}
\hat{\mathbf{D}}(\mathbf{\gamma}P)=\prod_n^{\otimes} \hat{D}_n(\gamma_n P)=\prod_n^{\otimes} \exp(\gamma_n P \left(a_n^\dagger-a_n\right)),
\end{equation}
where $\gamma_n$ are real parameters (with dimension 1/momentum), to be specified later. Note that the hat over $D$ reminds us that $P$ in its argument is an operator on the state space of the electron. With $P^\dagger =P,$ the operator $\hat{\mathbf{D}}(\mathbf{\gamma}P)$ is unitary, and $\hat{\mathbf{D}}^\dagger(\mathbf{\gamma}P)=\hat{\mathbf{D}}(-\mathbf{\gamma}P).$ As one can check,
\begin{equation}
\hat{D}^\dagger(\gamma_n P) a_n \hat{D}(\gamma_n P)=a_n + \gamma_n P, \ \  \hat{D}^\dagger(\gamma_n P) a_n^\dagger a_n \hat{D}(\gamma_n P)=a_n^\dagger a_n + \gamma_n P\left(a_n^\dagger + a_n \right) + \gamma_n^2 P^2.
\end{equation}

Now let us choose
\begin{equation}
\gamma_n=\frac{e\mathcal{A}_n}{m\hbar \omega_n}. \label{gammas}
\end{equation}
This leads to
\begin{eqnarray}\label{Htilde}
\widetilde{H}=\hat{\mathbf{D}}^\dagger(\mathbf{\gamma} P) H  \hat{\mathbf{D}}(\mathbf{\gamma} P)&=& \frac{P^2}{2m(\mathbf{\gamma})} + \sum_n \hbar \omega_n\left( a_n^\dagger a_n + \frac{1}{2}\right) \nonumber \\
&=&H_e^\prime\otimes\mathbf{1}+\mathbf{1}\otimes H_f,
\end{eqnarray}
where we used that $\left[\hat{\mathbf{D}}(\mathbf{\gamma} P), P\right]=0$ and introduced the notation 
\begin{equation}
1/m(\mathbf{\gamma})=1/m-2\sum_n \hbar \omega_n \gamma_n^2.
\end{equation}
In the following, the real parameters $\gamma_n$ will be fixed as given by Eq.~(\ref{gammas}). Note that under usual circumstances, for momenta $p$ that correspond to nonrelativistic velocities, we have $|\gamma_n p|\ll 1.$

Clearly, $\widetilde{H}$ given by Eq.~(\ref{Htilde}) is diagonal in the tensor product basis of the eigenstates of the operators $P$ and $N_n=a^\dagger_n a_n$. That is, by using the notation
\begin{equation}
\vert\mathbf{n}\rangle=\prod_i^{\otimes}|n_i\rangle,
\end{equation}
(where $N_i\vert n_i\rangle=n_i \vert n_i\rangle$), we can write
\begin{equation}\label{Htilde_eigen}
\widetilde{H} \vert p\rangle |\mathbf{n}\rangle= \left [ \frac{p^2}{2m(\boldsymbol{\gamma})} + \sum_i \hbar \omega_i\left( n_i+ \frac{1}{2}\right) \right ]\vert p\rangle |\mathbf{n}\rangle =E(p,\mathbf{n}) \vert p\rangle |\mathbf{n}\rangle.
\end{equation}

Using the unitarity of $\hat{\mathbf{D}},$ we can rewrite Eq.~(\ref{Htilde}) as
$\hat{\mathbf{D}}(\mathbf{\gamma} P)\widetilde{H}=H\hat{\mathbf{D}}(\mathbf{\gamma} P).$
In this way, $H$ can be diagonalized:
\begin{equation}
H \hat{\mathbf{D}}(\mathbf{\gamma} P)\vert p\rangle |\mathbf{n}\rangle=E(p,\mathbf{n})\hat{\mathbf{D}}(\mathbf{\gamma} P) \vert p\rangle |\mathbf{n}\rangle.
\end{equation}

Let us return to the interpretation of $m(\boldsymbol{\gamma})$ at this point. It is important to recall that we are using a nonrelativistic model which is valid in the long-wavelength limit. Additionally, even with many, but a finite number of modes taken into account, we have $|m-m(\mathbf{\gamma})|\ll m.$ It is important to emphasize that although the change of the prefactor before the operator $P^2$ can formally be taken into account by the $m\rightarrow m(\mathbf{\gamma})$ substitution, the present nonrelativistic description is different from mass renormalization in the QED sense. As we have shown above, the eigenstates of $H$ are coupled matter-field states (the parameters of the displacements of the field states depend on the value of the momentum $p$). In that sense, the appearance of $m(\mathbf{\gamma})$ is analogous to the change of the eigenenergies for dressed atomic states in quantum optical models (see e.g.~\cite{WallsMilburn1994}). Because of the interaction, the eigenenergies of the complete system will not be simply the sum of the eigenenergies of the (free) material system and the (free) electromagnetic field. It is exactly the energy difference caused by the interaction that can be taken into account by replacing $m$ with $m(\mathbf{\gamma})$.

\subsection{Time evolution}
By expanding the time evolution operators $U(t)=\exp(-iHt/\hbar)$ and $\widetilde{U}(t)=\exp(-i\widetilde{H}t/\hbar)=U_e(t)U_f(t)$ in power series, Eq.~(\ref{Htilde}) leads to
\begin{equation}\label{Utilde}
\hat{\mathbf{D}}(\mathbf{\gamma} P) \widetilde{U}(t) = U(t) \hat{\mathbf{D}}(\mathbf{\gamma} P).
\end{equation}
That is, by using Eq.~(\ref{Htilde_eigen}),
\begin{equation}\label{Uevol}
U(t) \hat{\mathbf{D}}(\mathbf{\gamma} P) \vert p\rangle |\mathbf{n}\rangle= \exp\left(-i\frac{E(p,\mathbf{n})t}{\hbar}\right)\hat{\mathbf{D}}(\mathbf{\gamma} P) \vert p\rangle |\mathbf{n}\rangle,
\end{equation}
which means that by expanding the initial state on the basis of $\hat{\mathbf{D}}(\mathbf{\gamma} P) \vert p\rangle |\mathbf{n}\rangle$, the time evolution is straightforward. Eq.~(\ref{Uevol}) can be simplified further by using the fact that $P\vert p\rangle=p \vert p \rangle$ and by replacing $\hat{\mathbf{D}}(\mathbf{\gamma} P)$ by a tensor product of the usual displacement operators $\mathbf{D}(\boldsymbol{\gamma}p)$ (note the absence of the hat in the notation) that act on the field states only:
\begin{equation}\label{Dprod}
\mathbf{D}(\boldsymbol{\gamma}p)=\prod_n^{\otimes} D_n(\gamma_n p)=\prod_n^{\otimes} \exp(\gamma_n p \left(a_n^\dagger-a_n\right)).
\end{equation}

Considering a single mode only, we write explicitly:
\begin{equation}\label{nevol}
U(t)\int  \widetilde{\phi_e}(p) \vert p\rangle \sum_n c_{n} D(\gamma p)\vert n\rangle dp
=\int \exp\left(-i\frac{p^2 t}{2m(\boldsymbol{\gamma})\hbar}\right) \widetilde{\phi_e}(p) \vert p\rangle \sum_n e^{-i\omega(n+1/2)} c_{n} D(\gamma p)\vert n\rangle  dp.
\end{equation}

Eq.~(\ref{nevol}) can be easily generalized for the multimode case. Instead of writing double indices, let us focus on specific cases that are of special interest. Using a compact notation, the general result reads:
\begin{equation}\label{Ugeneral}
\vert\Phi(t)\rangle=U(t) \int \widetilde{\phi_e}(p) \vert p\rangle \vert \phi_f\rangle dp
=\int \left[\widetilde{\phi_e}(p) U_e^\prime(t) \vert p\rangle\right] \otimes \left[ \mathbf{D}(\mathbf{\gamma} p) U_f(t) \mathbf{D}^\dagger(\mathbf{\gamma} p) \vert \phi_f\rangle \right] dp.
\end{equation}

Considering coherent states, we can write
\begin{equation}\label{coheveol0}
U(t)\vert p \rangle\vert \boldsymbol{\alpha} \rangle
=U(t)\vert p \rangle \mathbf{D}(\boldsymbol{\gamma }p)\mathbf{D}^\dagger(\boldsymbol{\gamma}p) \mathbf{D}(\boldsymbol{\alpha})\vert \boldsymbol{0} \rangle
=\mathbf{D}(\mathbf{\gamma} p) \widetilde{U}(t) \vert p \rangle \mathbf{D}^\dagger(\boldsymbol{\gamma}p) \mathbf{D}(\boldsymbol{\alpha})\vert \boldsymbol{0} \rangle,
\end{equation}
and use the multiplication rule for displacement operators together with the well-known free time evolution of the coherent states. This allows us to obtain finally:
\begin{equation}\label{coheveol}
U(t)\int \widetilde{\phi_e}(p) \vert p\rangle\vert \boldsymbol{\alpha} \rangle dp
=\int \widetilde{\phi_e}(p) \exp\left(-i\frac{p^2 t}{2m(\boldsymbol{\gamma})\hbar}\right) \vert p\rangle \vert \boldsymbol{\alpha}(p,t) \rangle dp.
\end{equation}
Here
\begin{equation}\label{alphas}
|\alpha_n(p,t)\rangle
=|p\gamma_n+(\alpha_n-p\gamma_n) e^{-i\omega_n t} \rangle e^{i\delta_n(p,t)}, \ \ \
\delta_n(p,t)=p\gamma_n \mathrm{Im} \left(\alpha_n-(\alpha_n-p\gamma_n)e^{-i\omega_n t} \right).
\end{equation}
Note that -- in order to simplify the notation -- the states $|\alpha_n(p,t)\rangle$ include a time-dependent phase, $e^{i\delta_n(p,t)}.$

A similar calculation can be carried out for squeezed coherent states,
\begin{equation}\label{sqdef}
\vert \boldsymbol{\alpha}, \boldsymbol{z} \rangle=\mathbf{D}(\boldsymbol{\alpha})\mathbf{S}(\boldsymbol{z})|\mathbf{0}\rangle,
\end{equation}
where the multimode squeezing operator is given by
\begin{equation}\label{Sdef}
\mathbf{S}(\boldsymbol{z})=\prod_n^{\otimes} S_n(z_n)
=\prod_n^{\otimes} \exp(\frac{1}{2}\left(z_n (a_n^\dagger)^2-z_n^* a_n^2\right)).
\end{equation}
The result is
\begin{equation}\label{sqeveol}
U(t)\int \widetilde{\phi_e}(p) \vert p\rangle\vert \boldsymbol{\alpha}, \boldsymbol{z} \rangle dp
=\int \widetilde{\phi_e}(p) \exp\left(-i\frac{p^2 t}{2m(\boldsymbol{\gamma})\hbar}\right) \vert p\rangle \vert \boldsymbol{\alpha}(p,t), \boldsymbol{z}(t) \rangle dp,
\end{equation}
with $\alpha_n(p,t)$ and $\delta_n(p,t)$ being the same as in Eq.~(\ref{alphas}), and $z_n(t)=z_n(0)\exp(-2i\omega_n t).$

\section{Position expectation value and uncertainty}
The general framework above admits multiple viewpoints, among them a phase-space description \cite{Czirjak1996,Schleich2001}. Here we concentrate on the expectation value and uncertainty of the position operator.

To set the stage, let us first recall the classical dynamics governed by the velocity-gauge Hamiltonian
\begin{equation}
\mathcal{H}(t)=\frac{p^2}{2m}-\frac{epA_{cl}(t)}{m}.
\end{equation}
As one can check easily, we have
\begin{equation}\label{classical}
p(t)=p_0, \  \ \ x(t)=x_0+p_0t/m-e \overline{A}_{cl}(t)/m,
\end{equation}
where the constants $p_0$ and $x_0$ should be determined from the initial conditions (at $t=0$), and the notation
\begin{equation}
\overline{A_{cl}}(t)= \int_{0}^{t} A_{cl}(t') dt'
\end{equation}
was introduced. Note that $p$ above stands for the canonical momentum; it is a constant of motion, but the kinetic momentum (that is directly related to the velocity of the particle) clearly has time dependence: $p_{kin}(t)=p_0-eA_{cl}(t).$

Similarly, in the quantum mechanical case, the expectation value (as well as the uncertainty) of $P$ is a constant of motion, simply because $[H,P]=0.$ This can be seen directly as well, by using Eq.~(\ref{Ugeneral}):
\begin{eqnarray}\label{Pexp}
\langle P\rangle(t) \! \! \! \! &=&\! \! \! \! \int \! \! \! \int \widetilde{\phi_e}(p) \widetilde{\phi_e^*}(p^\prime) \langle p^\prime\vert U_e^{\dagger}(t) P U_e(t) \vert p\rangle  \langle \phi_f \vert \mathbf{D}(\mathbf{\gamma} p^\prime) U^{\dagger}_f(t) \mathbf{D}^\dagger(\mathbf{\gamma} p^\prime) \mathbf{D}(\mathbf{\gamma} p) U_f(t) \mathbf{D}^\dagger(\mathbf{\gamma} p) \vert \phi_f\rangle  dp dp^\prime \nonumber \\
\! \!  \! \! &=&\! \! \! \! \int \left| \widetilde{\phi_e}(p)\right|^2 p \langle \phi_f \vert \phi_f\rangle  dp =p_0,
\end{eqnarray}
where we used the identity $\langle p^\prime \vert P\vert p\rangle=p\delta(p-p^\prime),$ the fact that the evolution operators as well as the generalized displacements are unitary, and the normalization of the initial states.

For the direct calculation of $\langle X\rangle(t)$ we need the matrix element
\begin{equation}\label{delta}
\langle p\vert X\vert p^\prime\rangle=i \hbar \delta^\prime(p-p^\prime).
\end{equation}
In order to simplify the notation, let us use $\widetilde{\phi_e}(p,t)=\widetilde{\phi_e}(p)e^{-i\frac{p^2 t}{2m(\mathbf{\gamma})\hbar}}$ and introduce
\begin{equation}\label{Fp1p2}
F(p_1,p_2,t)=\widetilde{\phi_e}(p_1,t)\widetilde{\phi_e}^*(p_2,t) \langle \phi_f\vert \mathbf{D}(\mathbf{\gamma} p_2) U^\dagger_f(t) \mathbf{D}^\dagger(\mathbf{\gamma} p_2) \mathbf{D}(\mathbf{\gamma} p_1) U_f(t) \mathbf{D}^\dagger(\mathbf{\gamma} p_1) \vert \phi_f\rangle,
\end{equation}
which naturally appears in the integrands when expectation values are calculated. (Note that $F(p,p,t)=\left| \widetilde{\phi_e}(p)\right|^2.$)
With this notation, and using Eq.~(\ref{delta}), we obtain
\begin{equation}\label{Xexpderiv}
\langle X\rangle(t) = \frac{i\hbar}{2}\int \left .\frac{\partial}{\partial p_1}F(p_1,p_2, t)\right|_{p_1=p_2=p}-  \left. \frac{\partial}{\partial p_2}F(p_1,p_2,t)\right|_{p_1=p_2=p} dp.
\end{equation}

In order to proceed, the inner product appearing in $F(p_1,p_2,t)$ should be calculated. For example, for an initial tensor product of coherent states, we have
\begin{equation}\label{Xexpalpha}
F(p_1,p_2,t)=\widetilde{\phi_e}(p_1,t)\widetilde{\phi_e}^*(p_2,t)  \langle \boldsymbol{\alpha}(p_2,t) \vert \boldsymbol{\alpha}(p_1,t) \rangle,
\end{equation}
and we can use the well-known formula for $\langle \alpha|\beta\rangle,$ perform the derivation, and finally integrate over $p$ to obtain $\langle X\rangle (t).$ An analogous method can be applied to the case of squeezed coherent states as well. The details of the completely general calculation can be found in the Appendix.

On using the notation
\begin{equation}\label{Abar}
\overline{A_0}(t)= \int_{0}^{t} A_0^H(t') dt'=\sum_n i \frac{\mathcal{A}_n}{\omega_n} \left[(a_n e^{-i\omega_n t}-a^\dagger_n e^{i\omega_n t})-(a_n-a^\dagger_n)\right]
\end{equation}
for the time integral of the Heisenberg picture free vector potential operator, the general result can be formulated as
\begin{equation}\label{Xexpfinal}
\langle X\rangle(t) = \langle X\rangle(0) + \frac{p_0}{m(\boldsymbol{\gamma})}t-\frac{e\langle\overline{A}_0\rangle(t)}{m}+2\hbar p_0\sum_n \gamma_n^2 \sin\omega_n t.
\end{equation}
This expression -- apart from terms proportional to $\gamma^2_n$ -- is consistent with the classical dynamics (\ref{classical}). Let us emphasize that Eq.~(\ref{Xexpfinal}) is valid for all initial states given by Eq.~(\ref{initial}), with the last term being independent of the initial state of the field -- it essentially stems from the commutator of $a_n$ and $a^\dagger_n.$

\bigskip

For the calculation of the position uncertainty, it is possible to use
\begin{equation}\label{delta2}
\langle p\vert X^2 \vert p^{\prime}\rangle=-\hbar^2 \delta^{\prime\prime}(p-p^\prime).
\end{equation}
However, in this case it is considerably simpler to work in the Heisenberg picture, where, for any operator $B^H,$ we have
\begin{equation}\label{X2H}
B^H(t)=U^\dagger(t) B U(t)=\hat{\mathbf{D}}(\mathbf{\gamma} P) \widetilde{U}^\dagger(t) \hat{\mathbf{D}}(\mathbf{-\gamma}P) B \hat{\mathbf{D}}(\mathbf{\gamma} P) \widetilde{U}(t) \hat{\mathbf{D}}(-\mathbf{\gamma}P).
\end{equation}
Note that the Schr\"{o}dinger and Heisenberg pictures coincide at $t=0,$ i.e., $B^H(0)=B.$ As we see, $B$ is sandwiched between exponential operators. For the most important cases, the corresponding series expansion terminates, leading to concise results. E.g.,
\begin{equation}\label{XH}
X(t)=\underbrace{X(0)+P \left(\frac{t}{m(\boldsymbol{\gamma})}+2 \hbar \sum_n \gamma_n^2 \sin\omega_n t \right)}_{\mathfrak{X}(t)} - \underbrace{i\hbar \sum_n \left(\Gamma^*_n(t)a_n^\dagger-\Gamma_n(t) a_n\right)}_{\frac{e}{m}\overline{A}_0(t)} ,
\end{equation}
where we dropped the explicit notation of the Heisenberg picture, and introduced the abbreviation
\begin{equation}\label{Gamma}
\Gamma_n(t) = \gamma_n (1 - e^{-i \omega_n t}).
\end{equation}
This result is in accord with Eq.~(\ref{Xexpfinal}), and the separation
turns out to be very useful, since $\mathfrak{X}=\mathfrak{X}\otimes\mathbf{1}$ and $\overline{A}_0=\mathbf{1}\otimes \overline{A}_0.$ With this notation, one obtains
\begin{equation}\label{X2H}
X^2(t)=\mathfrak{X}^2(t) -2\frac{e}{m} \mathfrak{X}(t) \overline{A}_0(t) + \frac{e^2}{m^2} \overline{A}_0^2 (t).
\end{equation}
Therefore, the uncertainty of the field operator $\overline{A}_0(t)$ appears explicitly in the uncertainty of the position:
\begin{equation}\label{deltaX2}
\Delta X(t)=\sqrt{\langle X^2(t) \rangle -\langle X(t)\rangle^2} = \sqrt{\Delta^2\mathfrak{X}(t) + \frac{e^2}{m^2}\Delta^2 \overline{A}_0(t)}.
\end{equation}

Note that since $\mathfrak{X}$ already contains a contribution from the field, under special
circumstances, especially at the beginning of the time evolution, $\Delta X$ can be smaller than its value without the interaction. This possibility has already been pointed out in Ref.~\cite{Varro_NJP_2008}, and was also found in the context of HHG \cite{EvenTzur2024_BSVmotion}. However, without a cavity or the presence of additional bodies, the spreading of the electron wave function gets faster when it interacts with e.g.\ laser light.

It is instructive to regroup the terms in Eq.~(\ref{deltaX2}) by separating the field-independent universal contributions:
\begin{align}\label{deltaX22}
 &\Delta^2 X(t)=\Delta^2 X_{free}(t)+ \frac{e^2}{m^2}\Delta^2 \overline{A}_0(t) +4\hbar \Delta^2 P \frac{t}{m}\sum_n \gamma_n^2  \sin{\omega_n t}  \\
  &+ 4\hbar^2 \Delta^2 P \left( \sum_n \gamma_n^2\sin{\omega_n t}\right )^2
  + 2\hbar  \Big( \langle PX(0) + X(0)P\rangle - 2\langle X(0) \rangle \langle P \rangle \Big) \sum_n \gamma_n^2 \sin{\omega_n t}. \nonumber
\end{align}
Here, the first term describes the spreading of the wave packet without interaction, but with the $m\rightarrow m(\mathbf{\gamma})$ substitution (``free" time evolution). $\Delta^2 \overline{A}_0$ is clearly determined by the initial state of the field, while the remaining terms contain (generally weak) contributions that are the same for all field states.

Using Eqs.~(\ref{deltaX2}) and (\ref{deltaX22}), it can be calculated to what extent $\Delta X$ depends on the state of the field it interacts with. The analysis of the resulting analytic expressions allows a transparent physical interpretation. Specific examples will be given in the next section.

\section{Examples}
\subsection{Wave packet in single-mode fields}
For the sake of simplicity, we start with the case of single-mode fields (and omit the mode index). Note that since for tensor product initial states (\ref{initial},\ref{initialfield}) the expectation value of operator products acting on the state space of different modes factorizes, the generalization of the current results to the multimode case is trivial, see the next subsection. In the following, the initial state of the electron is assumed to be given by a normalized Gaussian centered at $x=0$:
\begin{equation}\label{Gaussian}
  \phi_e(x) = \left(\frac{1}{2\pi\sigma_x^2}\right)^{1/4} \exp\!\left[ -\frac{x^2}{4\sigma_x^2} +\frac{i}{\hbar}p_0 x \right], \ \ \ \widetilde{\phi}_e(p) = \left(\frac{2\sigma_x^2}{\pi\hbar^2}\right)^{1/4} \exp\!\left[ -\frac{\sigma_x^2}{\hbar^2}(p-p_0)^2 \right].
\end{equation}

\bigskip

Coherent states are eigenstates of the annihilation operator, so we can use the identities $\langle \alpha |a|\alpha\rangle=\alpha$ and $\langle \alpha |a^\dagger|\alpha\rangle=\alpha^*$ to calculate the expectation values appearing in Eqs.~(\ref{Xexpfinal}) and (\ref{deltaX2}). This leads to
\begin{equation}\label{Xexpcohexplicit}
\langle X\rangle(\alpha, t)  = \frac{p_0}{m(\boldsymbol{\gamma})}t-2\hbar \mathrm{Im}\left(\Gamma(t) \alpha \right)+2\hbar p_0 \gamma^2 \sin\omega t,
\end{equation}
and
\begin{equation}\label{deltaXalpha}
 \Delta^2 \overline{A}_0(\alpha,t)=\frac{\hbar^2 m^2}{e^2} \left|\Gamma(t)\right |^2.
\end{equation}
It is worth recalling that $\hbar \gamma= \frac{e\mathcal A}{m\omega}.$ As we can see, $\langle X\rangle(t) $  depends on the parameter $\alpha,$ the amplitude of the wave packet oscillations increases for larger values of $|\alpha|.$ On the other hand, $\alpha$ does not appear in the expression for $\Delta^2 X(t),$ i.e., for a driving that most closely resembles a classical excitation, the amplitude of the field does not play a role in the spreading of the electron's wave packet.

\bigskip

For squeezed initial states, $|\phi_f\rangle=|\alpha,z\rangle,$ the expectation value of the position turns out to be the same as without squeezing:
\begin{equation}\label{Xexpcohexplicit}
\langle X\rangle(\alpha,z, t)  = \langle X\rangle(\alpha, t),
\end{equation}
which is a direct consequence of the fact that $\langle \alpha, z| a|\alpha, z\rangle=\alpha.$ On the other hand, for the field-dependent term that appears in the position uncertainty, we obtain:
\begin{equation}\label{deltaXalphaz}
 \Delta^2 \overline{A}_0(\alpha,z,t)=\frac{\hbar^2 m^2}{e^2} \left|\cosh(r)\Gamma^*(t)-\sinh(r)e^{i\theta}\Gamma(t)\right |^2,
\end{equation}
where the parametrization $z=re^{i \theta}$ was used.

\begin{figure}
\centering
\begin{minipage}{0.45\textwidth}
  \centering
  \includegraphics[width=\linewidth]{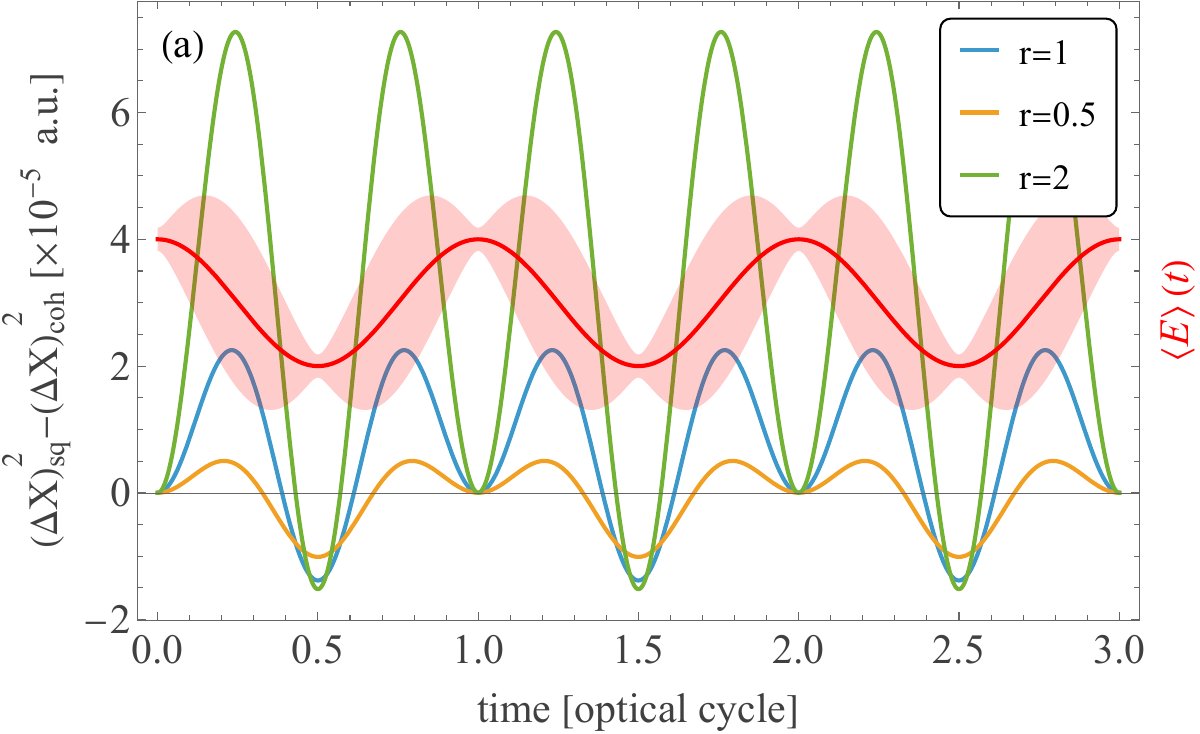}
\end{minipage}
\hfill
\begin{minipage}{0.45\textwidth}
  \centering
  \includegraphics[width=\linewidth]{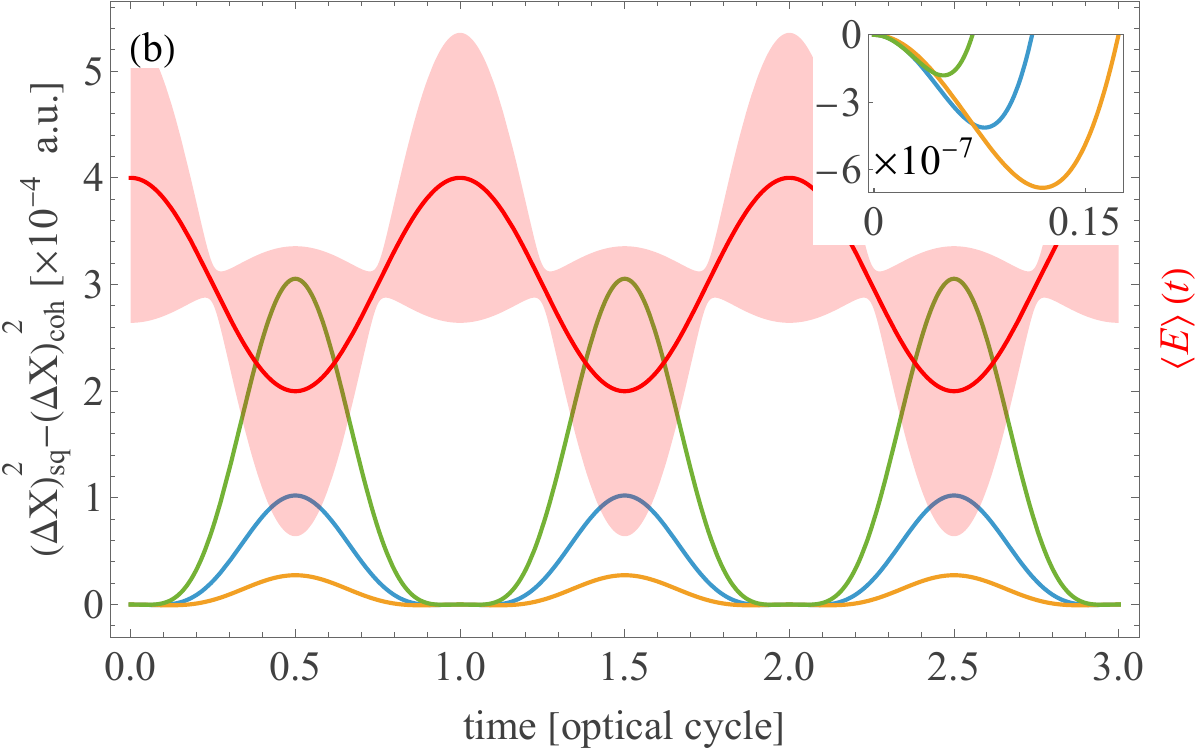}
\end{minipage}
\caption{Difference between the electron coordinate variance in a squeezed coherent field and in the corresponding coherent field.
(a) Amplitude squeezing ($\theta=0$), (b) phase squeezing ($\theta=\pi$), for the squeezing parameters indicated in panel (a).
The electron is initially prepared in the Gaussian state given by Eq.~(\ref{Gaussian}).
Negative values correspond to a reduction of the variance relative to the coherent-state case.
The coupling strength is $\gamma=0.002$~atomic units.}
\label{fig:single_mode_squeezed}
\end{figure}

Figure~\ref{fig:single_mode_squeezed} shows the difference of the position variance $\Delta^2 X$ calculated for a squeezed coherent field and the corresponding coherent field, for an electron initially prepared in the Gaussian state given by Eq.~(\ref{Gaussian}).
Panels (a) and (b) correspond to amplitude-squeezed and phase-squeezed fields (i.e., $\theta=0$ and $\theta=\pi$), respectively, for the squeezing parameters indicated in panel (a).
As can be seen, the coordinate variance in a squeezed state can become smaller than in the coherent case, and this behavior repeats in every optical cycle.
This feature follows directly from the analytical expression for the variance.
Specifically, the variance in the squeezed state is reduced relative to the coherent case at those times $t$ for which
\begin{equation}\label{ineqality}
\tanh(r)+\cos(\omega t-\theta)<0.
\end{equation}
Importantly, this condition is independent of both the coupling constant $\gamma$ and the coherent amplitude $\alpha$; it depends only on the squeezing parameters.
Therefore, for any squeezed coherent state (including squeezed vacuum), the variance is reduced within each optical cycle in the time interval
\begin{equation}\label{timeinterval}
t \in \left(\frac{\theta+\pi-\arccos(\tanh r)}{\omega},\,\frac{\theta+\pi+\arccos(\tanh r)}{\omega}\right).
\end{equation}
The magnitude of the effect, however, depends on the squeezing angle, as can be seen from the comparison of panels (a) and (b).

\bigskip

When the initial state is a Fock state, $|\phi_f\rangle=|n\rangle,$ the expectation values for field operators that are linear in $a$ or $a^\dagger$ vanish, e.g., $\langle \overline{A}_0\rangle(t)=0.$ Considering the uncertainty, we readily obtain:
\begin{equation}\label{XeFock}
\Delta^2 \overline{A}_0(n,t)=\frac{\hbar^2 m^2}{e^2} (2n +1) \left|\Gamma(t)\right |^2.
\end{equation}
This result can be generalized to the case of thermal states, which are incoherent sums of Fock states. In order to be more precise, we have to extend our description to the case of mixed quantum mechanical states that are described by density matrices. In particular, for a single mode thermal state at temperature $T$ we have:
\begin{equation}\label{rhoT}
\rho_f(T)=\frac{1}{Z}\sum_{n=0}^{\infty} e^{-\frac{\hbar \omega(n+1/2)}{kT}}|n\rangle\langle n|,
\end{equation}
where $k$ denotes Boltzmann's constant and $Z=\frac{\exp(-\hbar\omega/2kT)}{1-\exp(-\hbar\omega/kT)}.$ The initial state that is analogous to (\ref{initial}) is
\begin{equation}\label{rhoini}
\rho(T)=\rho_e \otimes \rho_f(T)=|\phi_e\rangle\langle \phi_e | \otimes \rho_f(T),
\end{equation}
where $|\phi_e\rangle$ corresponds to a Gaussian wave packet in coordinate representation, see Eq.~(\ref{Gaussian}). As one can check easily, $\langle X \rangle (T,t)=\mathrm{Tr}\left[ U(t) \rho(T) U^\dagger (t) X \right]$ still does not contain field-dependent terms, and
\begin{equation}\label{X2therm}
\Delta X(T,t)= \sqrt{\Delta^2\mathfrak{X}(t) + \hbar^2 \left|\Gamma(t)\right |^2 \coth\frac{\hbar \omega}{2kT}}.
\end{equation}

\begin{figure}
\centering
\begin{minipage}{0.45\textwidth}
  \centering
  \includegraphics[width=\linewidth]{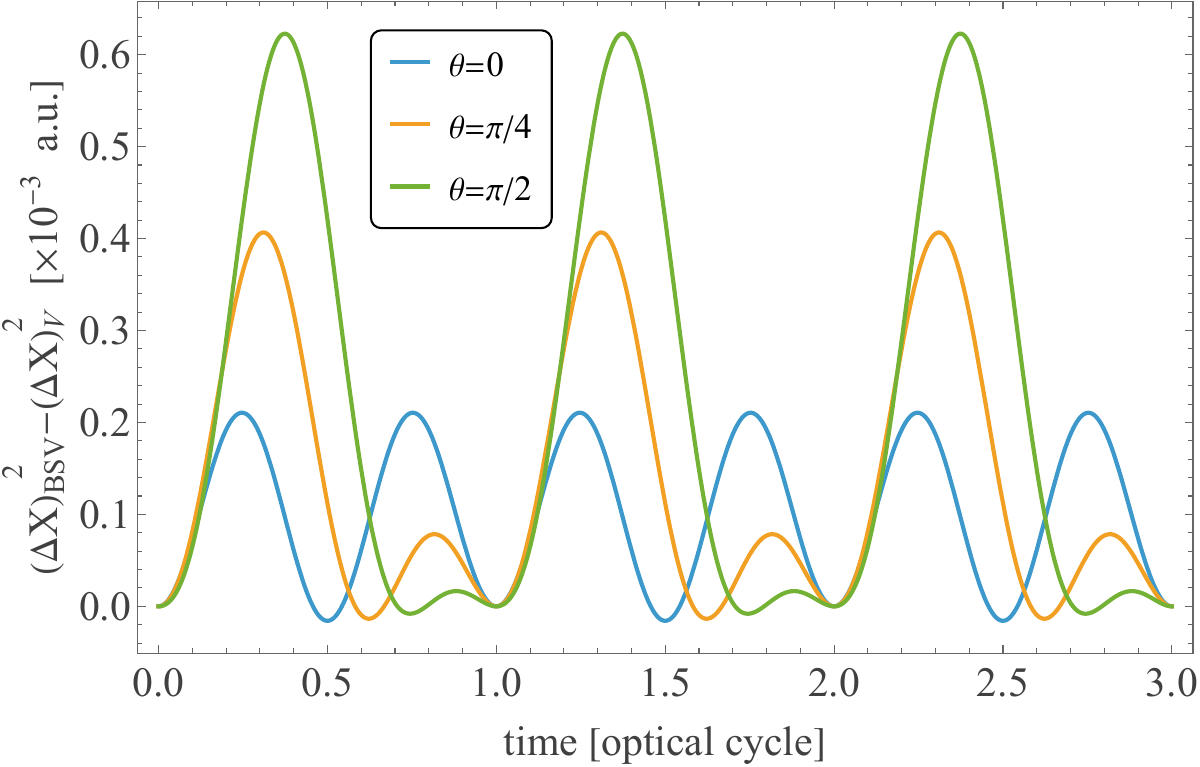}
\end{minipage}
\hfill
\begin{minipage}{0.45\textwidth}
  \centering
  \includegraphics[width=\linewidth]{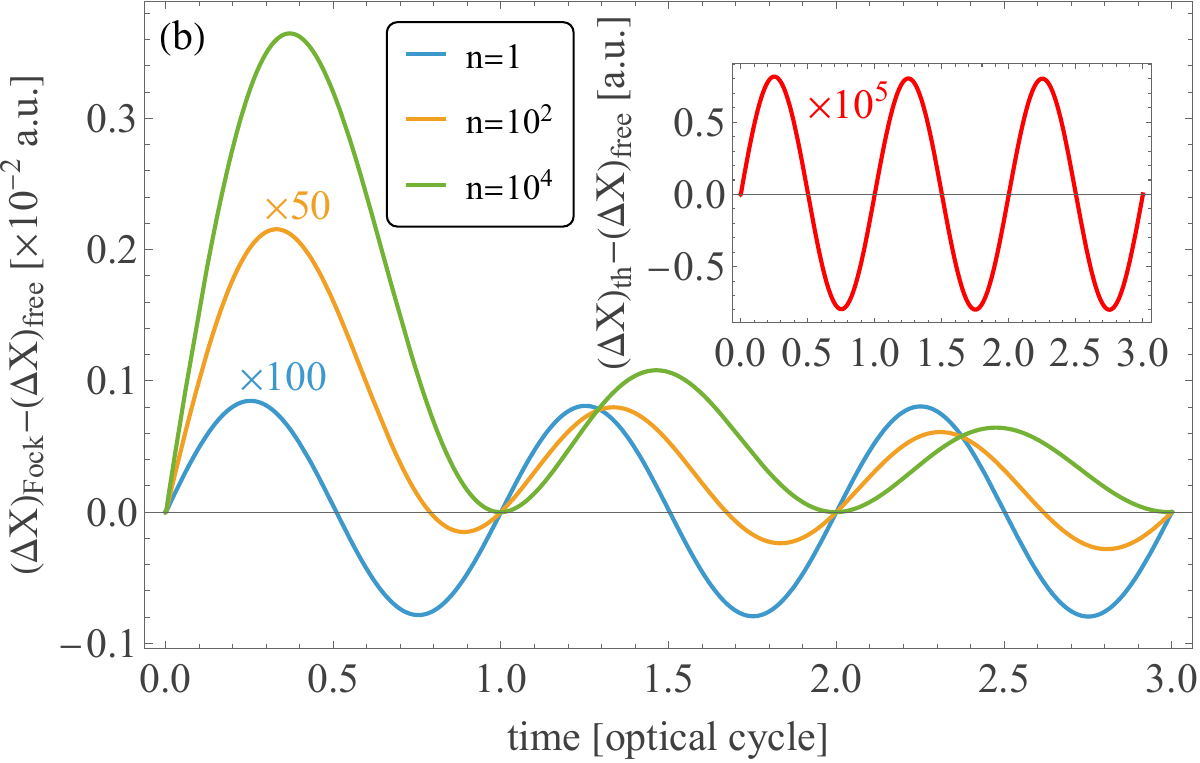}
\end{minipage}
\caption{Electron coordinate variance for fields with zero mean.
(a) Bright squeezed vacuum (BSV) with $r=2$ for different squeezing angles $\theta$; $\gamma=0.002$~a.u.
(b) Deviation of the electron coordinate variance from the free-electron case for Fock states with different photon numbers $n$.
The inset shows the corresponding result for a thermal state at $T=300$~K.}
\label{fig:single_mode_BSV_Fock}
\end{figure}

Figure~\ref{fig:single_mode_BSV_Fock} shows the electron coordinate variance for fields with zero mean: bright squeezed vacuum in panel (a), and Fock and thermal states in panel (b).
Since $\Delta X(t)$ does not depend on $\alpha$, the same result is obtained for squeezed coherent and squeezed vacuum states with identical squeezing parameters $(r,\theta)$.
This leads to the notable conclusion that the phase of the bright squeezed vacuum at the onset of the interaction is encoded in $\Delta X(t)$, as demonstrated in panel (a).
Panel (b) shows the deviation of the electron coordinate variance from the free-electron case for interaction with Fock states with different photon numbers.
Although the expectation value of the electric field vanishes, quantum fluctuations manifest themselves in the electron spreading and can lead to significant deviations from the free-particle case for large photon numbers.
The inset illustrates the corresponding result for a thermal state at a temperature of 300~K.

These results provide the physical intuition for the multimode case, where the interplay between different frequency components has to be taken into account as well.

\bigskip

\subsection{Wave packet in a multimode field}

The description of the mode structure of travelling electromagnetic waves, such as finite-duration laser pulses, requires a careful specification of the spatiotemporal degrees of freedom to be
quantized \cite{Blow1990,Sipe1995,Rohde2007,Fabre2020}.
In particular, the construction of a set of orthogonal field modes depends on the assumed
propagation geometry and on the temporal window over which the field is defined.
In the following, we adopt a simplified yet controlled description tailored to
laser-driven electron dynamics, which provides a transparent connection between classical
and quantum descriptions of the radiation field.

We consider an effective one-dimensional travelling-wave model in which all modes share a common
linear polarization and transverse spatial profile at the position of the particle.
Within this description, the classical electric field of a laser pulse is estimated on a
finite temporal interval by a Fourier series,
\begin{equation}
E_{\mathrm{cl}}(t)
\simeq
\sum_n
\left(
c_n e^{-i\omega_n t}
+
c_n^* e^{i\omega_n t}
\right),
\end{equation}
where the frequencies $\omega_n$ form a discrete set with spacing $\Delta\omega$.
The Fourier coefficients $c_n$ encode the spectral amplitude and phase of the pulse and are uniquely
determined by the choice of the temporal interval.

In the quantum description, the Fourier basis functions $e^{-i\omega_n t}$ are promoted to mode
functions of the quantized field.
To each such temporal mode, we associate bosonic annihilation and creation operators
$a_n$ and $a_n^\dagger$ obeying the usual commutation relation.
Let us recall that the Heisenberg picture electric field operator at the particle position is given by
\begin{equation}
{E_0^H}(t) =
i\sum_n \mathcal{E}_n
\left(
{a}_n e^{-i\omega_n t}
-
{a}_n^\dagger e^{i\omega_n t}
\right),
\end{equation}
where $\mathcal{E}_n = \mathcal{A}_n \omega_n$ encodes the effective quantization volume
associated with the chosen temporal and spatial mode basis.

The discretization of the frequency axis is tied to the introduction of a finite quantization time
$T_{\mathrm{box}}$, such that
\begin{equation}
\delta\omega = \frac{2\pi}{T_{\mathrm{box}}}.
\end{equation}
This relation is essential for the consistency of the quantized field description.
With this choice, the Fourier basis functions form an orthogonal set on the quantization
interval, allowing the field Hamiltonian to be expressed as a sum of independent harmonic oscillators.
Moreover, this choice ensures that the expectation value of the quantum-field energy reproduces the
classical energy of the laser pulse when the field is prepared in an appropriate multimode coherent
state.
The effective quantization volume entering $\mathcal{E}_n$ is therefore fixed by $T_{\mathrm{box}}$
and by the assumed transverse mode profile.
While $T_{\mathrm{box}}$ is an auxiliary parameter and must be taken larger than the pulse duration,
physical observables can depend on it unless the discrete mode set provides a sufficiently accurate
representation of the pulse spectrum.

The quantum state of the radiation field is assumed to factorize into a tensor product of single-mode states, as in Eq.~(\ref{initialfield}),
allowing for coherent, Fock, or squeezed states in each frequency mode.
For coherent and squeezed coherent fields, the expectation value of the electric field operator reproduces the corresponding classical field,
while quantum fluctuations are governed by the specific choice of the single-mode states $\vert \phi_{f,n}\rangle$.
Due to the tensor-product structure of the field state and the linear coupling to the field operators,
expectation values and variances decompose into mode-resolved contributions.
Consequently, the multimode generalization of the particle dynamics is obtained by summing the corresponding single-mode expressions over all contributing frequency modes, as shown in Eqs.~(\ref{XH}) and (\ref{deltaX22}).

In the case of squeezed coherent states, the reduction of the electron coordinate variance
relative to the unsqueezed case depends sensitively on the spectral composition of the field.
Due to the mode-resolved structure of the variance, this reduction is governed by a weighted
sum over frequency modes containing phase factors of the form $\omega_n t - \theta$.
As a consequence, the effect of squeezing is not determined solely by the squeezing parameter $r$,
but also by the spectral width of the pulse.
For a narrowband pulse centered at frequency $\omega$, one may obtain a transparent
sufficient condition by requiring that the dominant contributing modes lie within the
phase interval where $\tanh r + \cos(\omega_n t - \theta) < 0$.
Assuming a spectral support
$\omega_n \in [\omega - \Delta\omega/2,\, \omega + \Delta\omega/2]$,
this leads to the condition
\begin{equation}
\left|\omega t - \theta - \pi\right| + \frac{\Delta\omega}{2}\,t
< \arccos(\tanh r),
\end{equation}
which constrains the admissible spectral width $\Delta\omega$ of the pulse.
At the interaction time for which the central frequency satisfies
$\omega t - \theta = \pi$, the above condition simplifies to
\begin{equation}
\frac{\Delta\omega}{\omega}
<
\frac{2\,\arccos(\tanh r)}{\theta + \pi},
\end{equation}
where we have chosen the smallest positive solution for $t$.
This expression provides an upper bound on the relative spectral width for which
a reduction of the electron coordinate variance can be expected within this approximation.

We emphasize that this condition is sufficient but not necessary.
In general, partial cancellation between different frequency components may still lead to a
reduction of the variance even when the above bound is not strictly satisfied.
Nevertheless, the condition captures the essential requirement that the relevant spectral
components remain sufficiently phase-aligned around $\omega t - \theta \approx \pi$,
where the squeezing-induced noise suppression is most effective.
As the squeezing parameter $r$ increases, the allowed spectral width decreases,
reflecting the increased sensitivity of the effect to phase dispersion across the spectrum.

\begin{figure}
\centering
\begin{minipage}{0.45\textwidth}
  \centering
  \includegraphics[width=\linewidth]{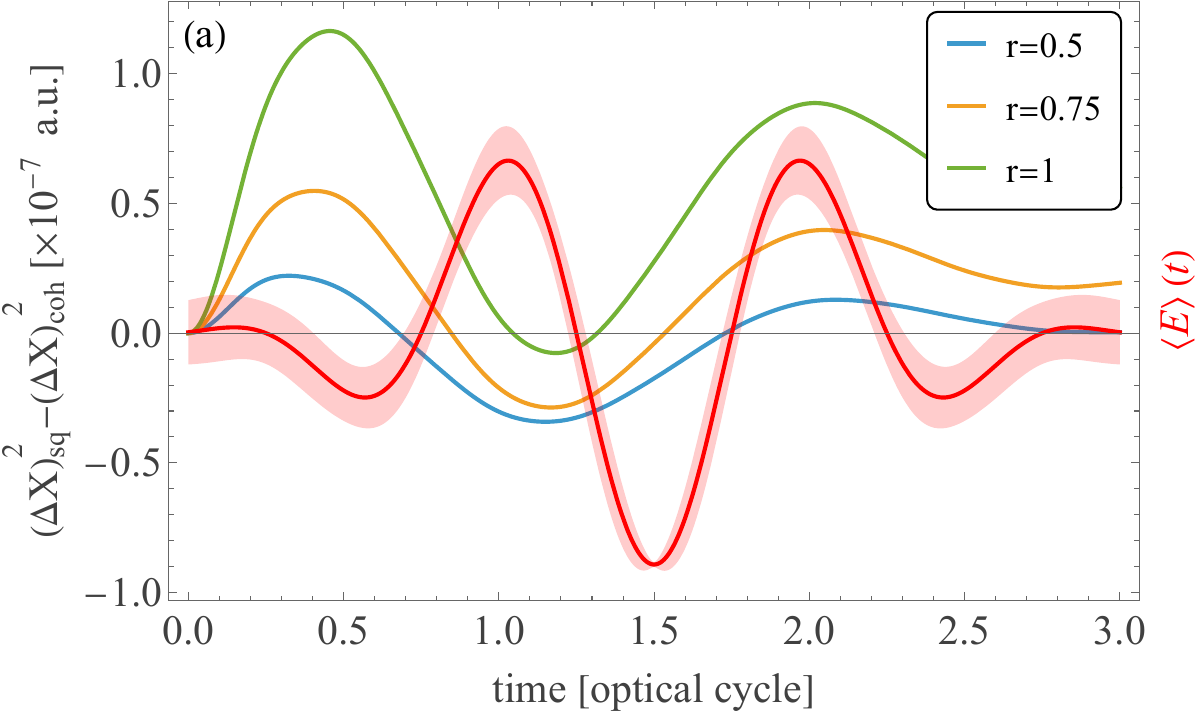}
\end{minipage}
\hfill
\begin{minipage}{0.45\textwidth}
  \centering
  \includegraphics[width=\linewidth]{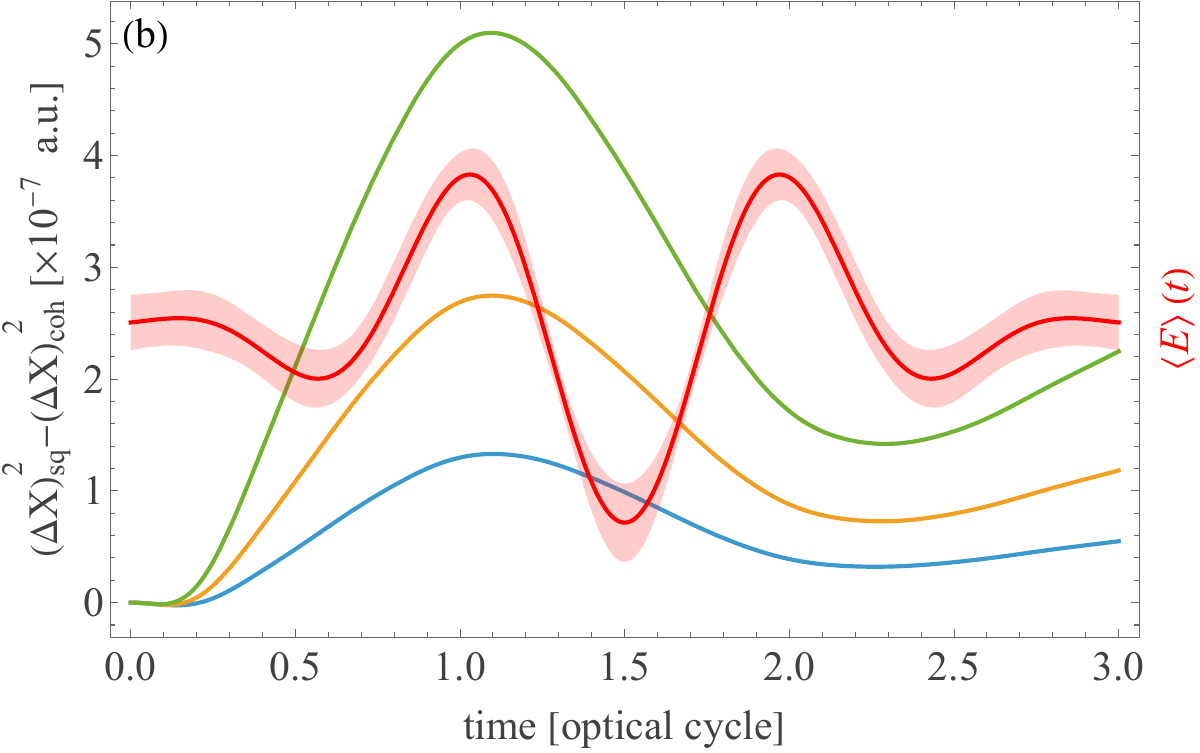}
\end{minipage}
\caption{Difference between the electron coordinate variance in a squeezed field and in the corresponding coherent field,
$(\Delta X)^2_{\mathrm{sq}} - (\Delta X)^2_{\mathrm{coh}}$, for amplitude-squeezed (a) and phase-squeezed (b) radiation and different values of the squeezing parameter $r$. Negative values correspond to a reduction of the electron coordinate variance relative to the coherent state case.
The driving laser pulse has a central wavelength of 1030~nm and a duration of three optical cycles.
The field is modeled using 400 discrete frequency modes with a transverse Gaussian profile and a $\sin^2$ temporal envelope, corresponding to a pulse energy of $1~\mu\mathrm{J}$.
The shaded background indicates the expectation value and variance of the electric field.}
\label{fig:multi_mode_squeezed}
\end{figure}

To illustrate these findings in a realistic setting, we present numerical results in Fig.~\ref{fig:multi_mode_squeezed}, where we show the difference between the electron coordinate variance in a squeezed field and in the corresponding coherent field,
$(\Delta X)^2_{\mathrm{sq}} - (\Delta X)^2_{\mathrm{coh}}$, as a function of time for different values of the squeezing parameter $r$.

In panel (a), amplitude-squeezed radiation is considered.
The variance reduction occurs only in limited temporal regions where the relevant frequency components satisfy the phase condition discussed above.
As $r$ increases, the magnitude of both variance reduction and enhancement grows, reflecting the stronger redistribution of quantum noise between the field quadratures.

In contrast, panel (b) shows the case of phase-squeezed radiation, where the variance is predominantly increased over the entire interaction time. This behavior reflects the intrinsic trade-off between conjugate field quadratures, whereby phase squeezing is accompanied by enhanced amplitude fluctuations, making it unfavorable for reducing the electron coordinate fluctuations in the present configuration.

In both cases, the temporal structure of the effect follows the oscillations of the driving field, shown in the background as the expectation value and variance of the electric field.
These results confirm that the reduction of electron fluctuations is governed not only by the squeezing strength but also by the spectral composition and phase coherence of the contributing modes.
These results highlight that, in realistic multimode fields, the effectiveness of squeezing-induced noise suppression is fundamentally limited by spectral phase dispersion.

\bigskip

\section{Summary}

We developed a fully analytic description of the dynamics of a free electron interacting with a multimode quantized electromagnetic field in the velocity gauge. By applying a momentum-dependent displacement transformation, the coupled Hamiltonian becomes diagonal in a product basis of electron momentum and photon number states, enabling closed-form expressions for the complete time evolution of arbitrary electron wave packets and field states. This approach yields explicit formulas for the position expectation value and spatial uncertainty, revealing a clear separation between universal contributions and terms that encode the quantum properties of the radiation field.

The expectation value of the position follows the classical trajectory up to small corrections proportional to the coupling parameters $\gamma_n^2$, and these corrections are largely insensitive to the quantum state of the field. In contrast, the wave packet spreading exhibits a pronounced dependence on the field state: the uncertainty of the radiation field is directly imprinted onto the electron's spatial uncertainty. This effect appears for coherent, Fock, squeezed, and bright squeezed vacuum states, and persists in multimode scenarios such as finite-duration laser pulses. The analytic expressions provide transparent physical insight into how quantum fluctuations of light influence electron motion, offering a foundation for interpreting strong-field experiments where field quantization plays a measurable role.
Our results could be relevant to well established strong-field scenarios and also new research directions, like electron microscopy combined with and enhanced by a cw laser \cite{Sun2023, Nabben2023}.

\section{Acknowledgment}

The work of B.G. Pusztai was supported by project TKP2021-NVA-09. 
Project no. TKP2021-NVA-09 has been implemented with the support 
provided by the Ministry of Innovation and Technology of Hungary from 
the National Research, Development and Innovation Fund, financed under 
the TKP2021-NVA funding scheme.
The ELI ALPS project (GINOP-2.3.6-15-2015-00001) is supported by the European Union and co-financed by the European Regional Development Fund.

\section{Appendix}
\subsection{Position expectation value}
Generally, in view of Eq.~(\ref{Xexpderiv}), we are to calculate the formal derivatives of $F(p_1,p_2,t)$:
\begin{eqnarray}\label{Xexpderiv2}
 \left .\frac{\partial}{\partial p_1}F(p_1,p_2, t)\right|_{p_1=p_2=p}&=& \widetilde{\phi_e^*}(p)\frac{\partial}{\partial \nonumber p}\widetilde{\phi_e}(p)- \left| \widetilde{\phi_e}(p)\right|^2 \frac{ip}{\hbar m}t  \\ &+& \left| \widetilde{\phi_e}(p)\right|^2 \sum_n \gamma_n \langle \phi_f\vert \mathbf{D}(\mathbf{\gamma} p) U^\dagger_e(t) \nonumber (a_n^\dagger-a_n) U_e(t) \mathbf{D}^\dagger(\mathbf{\gamma} p) \vert \phi_f\rangle \\ &-&
   \left| \widetilde{\phi_e}(p)\right|^2 \sum_n\langle \phi_f\vert (a_n^\dagger-a_n) \vert \phi_f\rangle.
\end{eqnarray}
Note that the derivation with respect to $p_2$ provides exactly the same result, but with opposite sign. Recalling Eq.~(\ref{Xexpderiv}), the first term on the right hand side is readily identified as the expectation value of the position operator in momentum representation, while the second one provides the second term in Eq.~(\ref{Xexpfinal}). The summand in the second line can be reformulated:
\begin{eqnarray}\label{Xexpderivpart}
\nonumber
\langle  \phi_f\vert \mathbf{D}(\mathbf{\gamma} p) U^\dagger_f(t) (a_n^\dagger-a_n) U_f(t) \mathbf{D}^\dagger(\mathbf{\gamma} p) \vert \phi_f\rangle &=& \langle \phi_f\vert \mathbf{D}(\mathbf{\gamma} p)  (a_n^\dagger e^{i\omega_n t} -a_n e^{-i\omega_n t})  \mathbf{D}^\dagger(\mathbf{\gamma} p) \vert \phi_f\rangle\\ &=&\langle \phi_f\vert (a_n^\dagger e^{i\omega_n t} -a_n e^{-i\omega_n t}) \vert \phi_f\rangle -2ip\gamma\sin\omega_n t.
\end{eqnarray}
Recalling Eq.~(\ref{Abar}), and combining the result above with the last term of Eq.~(\ref{Xexpderiv2}), we see that  $\langle \overline{A_0}\rangle(t)$ appears on the right hand side. Using that $F(p,p,t)=\left| \widetilde{\phi_e}(p)\right|^2,$ $\int | \widetilde{\phi_e}(p)|^2 dp=1$ and $\langle \phi_f|\phi_f\rangle=1,$ finally we obtain Eq.~(\ref{Xexpfinal}).

\bibliographystyle{apsrev4-2}
\bibliography{WPbib2}

@article{Pauli1938,
  author    = {W. Pauli and M. Fierz},
  title     = {Zur Theorie der Emission langwelliger Photonen},
  journal   = {Il Nuovo Cimento},
  volume    = {15},
  pages     = {167--188},
  year      = {1938},
  doi       = {10.1007/BF02959903}
}

@article{BlochNordsieck1937,
  author    = {Felix Bloch and Arnold Nordsieck},
  title     = {Note on the Radiation Field of the Electron},
  journal   = {Physical Review},
  volume    = {52},
  number    = {2},
  pages     = {54--59},
  year      = {1937},
  doi       = {10.1103/PhysRev.52.54}
}

@article{Gabor1950,
  author    = {Gabor, D.},
  title     = {Communication theory and physics},
  journal   = {Philosophical Magazine},
  volume    = {41},
  pages     = {1161--1187},
  year      = {1950}
}

@article{Sun2023,
  author    = {Sun, Feng-Xiao and Fang, Yiqi and He, Qiongyi and Liu, Yunquan},
  title     = {Generating optical cat states via quantum interference of multi-path free-electron--photons interactions},
  journal   = {Science Bulletin},
  volume    = {68},
  number    = {13},
  pages     = {1366--1371},
  year      = {2023},
  doi       = {10.1016/j.scib.2023.06.006}
}

@article{Nabben2023,
  author    = {Nabben, D. and Kuttruff, J. and Stolz, L. and others},
  title     = {Attosecond electron microscopy of sub-cycle optical dynamics},
  journal   = {Nature},
  volume    = {619},
  pages     = {63--67},
  year      = {2023},
  doi       = {10.1038/s41586-023-06074-9}
}

@article{Varro2010,
  author    = {Varro, S.},
  title     = {Entangled states and entropy remnants of a photon-electron system},
  journal   = {Physica Scripta},
  volume    = {T140},
  pages     = {014038},
  year      = {2010},
  doi       = {10.1088/0031-8949/2010/T140/014038}
}

@article{Smith1946,
  author    = {Smith, L. P.},
  title     = {Quantum effects in the interaction of electrons with high frequency fields and the transition to classical theory},
  journal   = {Physical Review},
  volume    = {69},
  pages     = {195--210},
  year      = {1946}
}

@book{Spohn2004,
  author    = {Herbert Spohn},
  title     = {Dynamics of Charged Particles and Their Radiation Field},
  publisher = {Cambridge University Press},
  year      = {2004},
  isbn      = {9780521836975},
  doi       = {10.1017/CBO9780511616882}
}

@book{WallsMilburn1994,
  author    = {D. F. Walls and G. J. Milburn},
  title     = {Quantum Optics},
  publisher = {Springer},
  year      = {1994},
  isbn      = {9783540572807},
  doi       = {10.1007/978-3-662-02520-4}
}

@book{CohenTannoudji1989,
  author    = {Claude Cohen-Tannoudji and Jacques Dupont-Roc and Gilbert Grynberg},
  title     = {Photons and Atoms: Introduction to Quantum Electrodynamics},
  publisher = {Wiley},
  year      = {1989},
  isbn      = {9780471815186}
}

@article{EvenTzur2024_BSVmotion,
  author    = {Matan Even Tzur and Oren Cohen},
  title     = {Motion of charged particles in bright squeezed vacuum},
  journal   = {Light: Science \& Applications},
  volume    = {13},
  pages     = {41},
  year      = {2024},
  doi       = {10.1038/s41377-024-01381-w}
}

@article{EvenTzur2024_HHGsqueezed,
  author    = {Matan Even Tzur and Michael Birk and Alexey Gorlach and Ido Kaminer and Michael Kr\"uger and Oren Cohen},
  title     = {Generation of squeezed high-order harmonics},
  journal   = {Physical Review Research},
  volume    = {6},
  pages     = {033079},
  year      = {2024},
  doi       = {10.1103/PhysRevResearch.6.033079}
}

@article{Krausz2024,
  author       = {Krausz, Ferenc},
  title        = {Attosecond Physics: Past, Present, Future},
  journal      = {Reviews of Modern Physics},
  volume       = {96},
  number       = {3},
  pages        = {030502},
  year         = {2024},
  doi          = {10.1103/RevModPhys.96.030502}
}

@article{Rasputnyi2024_BSVHHG,
  author    = {Andrei Rasputnyi and Zhaopin Chen and Michael Birk and Oren Cohen and Ido Kaminer and Michael Kr\"uger and Denis Seletskiy and Maria Chekhova and Francesco Tani},
  title     = {High Harmonic Generation by Bright Squeezed Vacuum},
  journal   = {arXiv preprint},
  eprint    = {2403.15337},
  archivePrefix = {arXiv},
  year      = {2024}
}

@article{Heimerl2025_BSVphotoemission,
  author    = {J. Heimerl and collaborators},
  title     = {Bright squeezed vacuum reveals hidden quantum effects in strong-field physics},
  journal   = {Nature Physics},
  year      = {2025},
  doi       = {10.1038/s41567-025-03087-1}
}

@article{Blow1990,
  author    = {K. J. Blow and R. Loudon and S. J. D. Phoenix and T. J. Shepherd},
  title     = {Continuum fields in quantum optics},
  journal   = {Physical Review A},
  volume    = {42},
  number    = {7},
  pages     = {4102--4114},
  year      = {1990},
  doi       = {10.1103/PhysRevA.42.4102}
}

@article{Rohde2007,
  author    = {P. P. Rohde and W. Mauerer and C. Silberhorn},
  title     = {Spectral structure and decompositions of optical states},
  journal   = {New Journal of Physics},
  volume    = {9},
  pages     = {91},
  year      = {2007},
  doi       = {10.1088/1367-2630/9/4/091}
}

@article{Fabre2020,
  author    = {Claude Fabre and Nicolas Treps},
  title     = {Modes and states in quantum optics},
  journal   = {Reviews of Modern Physics},
  volume    = {92},
  pages     = {035005},
  year      = {2020},
  doi       = {10.1103/RevModPhys.92.035005}
}

@article{Stammer2023_QEDHHG,
  author    = {Philipp Stammer and Javier Rivera-Dean and Andrew Maxwell and Theocharis Lamprou and Andr\'es Ord\'o\~nez and Marcelo F. Ciappina and Paraskevas Tzallas and Maciej Lewenstein},
  title     = {Quantum Electrodynamics of Intense Laser-Matter Interactions: A Tool for Quantum State Engineering},
  journal   = {PRX Quantum},
  volume    = {4},
  pages     = {010201},
  year      = {2023},
  doi       = {10.1103/PRXQuantum.4.010201}
}

@article{RiveraDean2024,
  author    = {Rivera-Dean, J. and Crispin, H. B. and Stammer, P. and Lamprou, Th. and Pisanty, E. and Kruger, M. and Tzallas, P. and Lewenstein, M. and Ciappina, M. F.},
  title     = {Squeezed states of light after high-order harmonic generation in excited atomic systems},
  journal   = {Physical Review A},
  volume    = {110},
  pages     = {063118},
  year      = {2024},
  doi       = {10.1103/PhysRevA.110.063118}
}

@article{Lewenstein2021,
  author    = {Lewenstein, M. and Ciappina, M. F. and Pisanty, E. and Rivera-Dean, J. and Stammer, P. and Lamprou, Th. and Tzallas, P.},
  title     = {Generation of optical Schrodinger cat states in intense laser-matter interactions},
  journal   = {Nature Physics},
  volume    = {17},
  pages     = {1104--1108},
  year      = {2021},
  doi       = {10.1038/s41567-021-01317-w}
}

@article{Theidel2024_QOptHHG,
  author    = {David Theidel and Viviane Cotte and Ren\'e Sondenheimer and Viktoriia Shiriaeva and Marie Froidevaux and Vladislav Severin and Adam Merdji-Larue and Philip Mosel and Sven Fr\"ohlich and collaborators},
  title     = {Evidence of the Quantum Optical Nature of High-Harmonic Generation},
  journal   = {PRX Quantum},
  volume    = {5},
  pages     = {040319},
  year      = {2024},
  doi       = {10.1103/PRXQuantum.5.040319}
}

@article{Theidel2025_DisplacedSqueezed,
  author    = {David Theidel and Viviane Cotte and Philipp Heinzel and Hugo Griguer and Maximilian Weis and Ren\'e Sondenheimer and others},
  title     = {Observation of a displaced squeezed state in high-harmonic generation},
  journal   = {Physical Review Research},
  volume    = {7},
  number    = {3},
  pages     = {033223},
  year      = {2025},
  doi       = {10.1103/PhysRevResearch.7.033223}
}

@article{Corkum1993_Threestep,
  author    = {Paul B. Corkum},
  title     = {Plasma perspective on strong field multiphoton ionization},
  journal   = {Physical Review Letters},
  volume    = {71},
  number    = {13},
  pages     = {1994--1997},
  year      = {1993},
  doi       = {10.1103/PhysRevLett.71.1994}
}

@article{Lewenstein1994_HHGmodel,
  author    = {Maciej Lewenstein and Ph. Balcou and M. Yu. Ivanov and Anne L'Huillier and Paul B. Corkum},
  title     = {Theory of high-harmonic generation by low-frequency laser fields},
  journal   = {Physical Review A},
  volume    = {49},
  number    = {3},
  pages     = {2117--2132},
  year      = {1994},
  doi       = {10.1103/PhysRevA.49.2117}
}

@article{BergouVarro1981_Nonrel,
  author    = {J\'anos Bergou and S\'andor Varr\'o},
  title     = {Nonrelativistic theory of nonlinear Compton scattering},
  journal   = {Journal of Physics A: Mathematical and General},
  volume    = {14},
  number    = {9},
  pages     = {2281--2296},
  year      = {1981},
  doi       = {10.1088/0305-4470/14/9/021}
}

@article{Varro2021_Photonics,
  author    = {S\'andor Varr\'o},
  title     = {Quantum optical aspects of high-order harmonic generation},
  journal   = {Photonics},
  volume    = {8},
  number    = {7},
  pages     = {269},
  year      = {2021},
  doi       = {10.3390/photonics8070269}
}

@article{FarkasToth1992_Attopulses,
  author    = {Gy\H{o}z\H{o} Farkas and Csaba T\'oth},
  title     = {Proposal for attosecond light pulse generation using laser induced multiple-harmonic conversion processes in rare gases},
  journal   = {Physics Letters A},
  volume    = {168},
  number    = {5-6},
  pages     = {447--450},
  year      = {1992},
  doi       = {10.1016/0375-9601(92)90534-S}
}

@article{Bogoliubov1947,
  author    = {Bogoliubov, N. N.},
  title     = {On the theory of superfluidity},
  journal   = {J. Phys. (USSR)},
  volume    = {11},
  pages     = {23--32},
  year      = {1947}
}

@article{Andrianov2024PRA,
  title   = {Formation of nonclassical and non-Gaussian states of a strong electromagnetic field due to its interaction with free electrons produced by ionization of a target gas},
  author  = {Andrianov, Evgeny S. and Tolstikhin, Oleg I.},
  journal = {Physical Review A},
  volume  = {110},
  number  = {2},
  pages   = {023115},
  year    = {2024},
  doi     = {10.1103/PhysRevA.110.023115},
  publisher = {American Physical Society}
}

@article{Andrianov2025PRA,
  title = {Formation of a Schr\"odinger cat state of a strong circularly polarized laser field due to Thomson scattering by free electrons},
  author = {Andrianov, Evgeny S. and Tolstikhin, Oleg I.},
  journal = {Phys. Rev. A},
  volume = {112},
  issue = {1},
  pages = {013117},
  numpages = {9},
  year = {2025},
  month = {Jul},
  publisher = {American Physical Society},
  doi = {10.1103/3gjp-f7br},
  url = {https://link.aps.org/doi/10.1103/3gjp-f7br}
}

@article{McPherson87,
author = {McPherson, A. and Gibson, G. and Jara, H. and Johann, U. and Luk, T. S. and McIntyre, I. A. and Boyer, K. and Rhodes, C. K.},
journal = {J. Opt. Soc. Am. B},
number = {4},
pages = {595--601},
publisher = {OSA},
title = {Studies of multiphoton production of vacuum-ultraviolet radiation in the rare gases},
volume = {4},
year = {1987}
}

@article{Ferray88,
  author={Ferray, M. and L'Huillier, A. and Li, X. F. and Lompre, L. A. and Mainfray, G. and Manus, C.},
  title={Multiple-harmonic conversion of 1064 nm radiation in rare gases},
  journal={J. Phys. B: At. Mol. Phys.},
  volume={21},
  number={3},
  pages={L31},
  year={1988}
}

@article{Tsatrafyllis2017,
Author = {Tsatrafyllis, N. and Kominis, I. K. and Gonoskov, I. A. and Tzallas, P.},
Title = {High-order harmonics measured by the photon statistics of the infrared
   driving-field exiting the atomic medium},
Journal = {Nat. Comm.},
Year = {2017},
Volume = {8},
Type = {Article},
Article-Number = {15170},
pages = {15170}
}

@article{T19,
  title = {Quantum Optical Signatures in a Strong Laser Pulse after Interaction with Semiconductors},
  author = {Tsatrafyllis, N. and K\"uhn, S. and Dumergue, M. and Foldi, P. and Kahaly, S. and Cormier, E. and Gonoskov, I. A. and Kiss, B. and Varju, K. and Varro, S. and Tzallas, P.},
  journal = {Phys. Rev. Lett.},
  volume = {122},
  issue = {19},
  pages = {193602},
  numpages = {6},
  year = {2019},
  publisher = {American Physical Society},
  doi = {10.1103/PhysRevLett.122.193602}
  }

@article{GCVF2016,
  title = {Quantum-optical model for the dynamics of high-order-harmonic generation},
  author = {Gombk\"ot\ifmmode \mbox{\H{o}}\else \H{o}\fi{}, \'{A} and Czirj\'ak, A. and Varr\'o, S. and F\"oldi, P.},
  journal = {Phys. Rev. A},
  volume = {94},
  issue = {1},
  pages = {013853},
  numpages = {8},
  year = {2016},
  publisher = {American Physical Society},
}

@article{GVMF20,
  title = {High-order harmonic generation as induced by a quantized field: Phase-space picture},
  author = {Gombk\"ot{\H{o}}, \'A and Varr\'o, S and Mati, P and P. F\"oldi},
  journal = {Phys. Rev. A},
  volume = {101},
  pages = {013418},
  numpages = {10},
  year = {2020}
}

@article{GVP24,
  title = {Parametric model for high-order harmonic generation with quantized fields},
  author = {Gombk\"ot{\H{o}}, \'Akos and Varr\'o, S\'andor and Pusztai, B\'ela G\'abor and Magashegyi, Istv\'an and Czirj\'ak, Attila and Hack, Szabolcs and F\"oldi, P\'eter},
  journal = {Phys. Rev. A},
  volume = {109},
  issue = {5},
  pages = {053717},
  numpages = {11},
  year = {2024},
  month = {May},
  publisher = {American Physical Society},
  doi = {10.1103/PhysRevA.109.053717},
  url = {https://link.aps.org/doi/10.1103/PhysRevA.109.053717}
}

@article{Varro_NJP_2008,
  author       = {S. Varro},
  title        = {Entangled photon-electron states and the number-phase minimum uncertainty states of the photon field},
  journal      = {New Journal of Physics},
  volume       = {10},
  pages        = {053028},
  year         = {2008},
  doi          = {10.1088/1367-2630/10/5/053028}
}

@article{Sipe1995,
  author    = {Sipe, J. E.},
  title     = {Photon wave functions},
  journal   = {Physical Review A},
  volume    = {52},
  number    = {3},
  pages     = {1875--1883},
  year      = {1995},
  doi       = {10.1103/PhysRevA.52.1875}
}

@article{Arend2025,
  author    = {Arend, Germaine and Huang, Guanhao and Feist, Armin and Yang, Yujia and Henke, Jan-Wilke and Qiu, Zheru and Jeng, Hao and Raja, Arslan Sajid and Haindl, Rudolf and Wang, Rui Ning and Kippenberg, Tobias J. and Ropers, Claus},
  title     = {Electrons herald non-classical light},
  journal   = {Nature Physics},
  year      = {2025},
  volume    = {21},
  pages     = {1855--1862},
  doi       = {10.1038/s41567-025-03033-1}
}

@article{Czirjak1996,
  author    = {Czirjak, A. and Benedict, M. G.},
  title     = {Joint Wigner function for atom-field interactions},
  journal   = {Quantum and Semiclassical Optics},
  volume    = {8},
  number    = {5},
  pages     = {975--981},
  year      = {1996},
  doi       = {10.1088/1355-5111/8/5/003}
}

@book{Schleich2001,
  author    = {Schleich, W. P.},
  title     = {Quantum Optics in Phase Space},
  publisher = {Wiley-VCH},
  address   = {Weinheim, Germany},
  year      = {2001},
  isbn      = {352729435X}
}

@article{Kern2026,
  author    = {Kern, Yuval and Nisim, Ido and Birk, Michael and Rasputnyi, Andrei and Behar, Doron and Chen, Zhaopin and Kaminer, Ido and Sidorenko, Pavel and Cohen, Oren and Krueger, Michael},
  title     = {Single-shot pulse retrieval of femtosecond bright squeezed vacuum},
  journal   = {Optica},
  volume    = {13},
  number    = {3},
  pages     = {395--399},
  year      = {2026},
  doi       = {10.1364/OPTICA.580767}
}

\end{document}